\documentclass[twocolumn,showpacs,showkeys,superscriptaddress]{revtex4}

\usepackage{amsmath}
\usepackage{bbold}
\usepackage{amsfonts}
\usepackage{amssymb}
\usepackage{pbsi}
\usepackage[T1]{fontenc}
\usepackage{xcolor}
\usepackage{graphicx}
\usepackage{subfig}
\usepackage{float}
\usepackage{color}
\usepackage{hyperref}
\hypersetup{colorlinks = true,
	linkcolor = red,
	anchorcolor = blue,
	citecolor = blue,
	filecolor = blue,
	urlcolor = blue}

\begin{document}

\title{Accelerating self-modulated nonlinear waves in weakly and strongly magnetized relativistic plasmas}

\author{Felipe A. Asenjo}
\email{felipe.asenjo@uai.cl}
\affiliation{Facultad de Ingenier\'ia y Ciencias,
Universidad Adolfo Ib\'a\~nez, Santiago 7491169, Chile.}

\date{\today}

\begin{abstract}
 It is known that a nonlinear Schr\"odinger equation describes the self-modulation of a large amplitude circularly polarized wave in relativistic electron-positron
  plasmas in the  weakly and strongly magnetized limits. 
 Here, we show that such equation can be written as a
 modified second Painlev\'e equation, producing accelerated propagating wave solutions for those nonlinear plasmas. This solution even allows the plasma wave to reverse its direction of propagation. The acceleration parameter depends on the plasma magnetization. This accelerating solution is different to the usual soliton solution propagating at constant speed.
 \end{abstract}



\maketitle

\section{Introduction}

One of the nonlinear effects present in relativistic hot magnetized electron-positron plasma is the self-modulation of  circularly polarized electromagnetic waves or Alfv\'en waves. The large-amplitude of the electromagnetic wave, and a background magnetic field, can  modify the relativistic motion of particles in a significant way. Thus, the self-modulation depends on the magnetization of the plasma. This nonlinear process has been thoroughly studied for the weakly magnetized plasma and strongly magnetized plasma cases in Refs. \cite{faz1} and \cite{faz2}, respectively.

Interestingly, the self-modulation gives origin to soliton plasma wave solutions, propagating at constant speed \cite{faz1,faz2}. However, there exit other kind of accelerating plasma wave solution that deserved to be explored for relativistic plasmas. This work is devoted to show that there are  nonlinear plasma waves that accelerate  due to the self-modulation of the magnetized plasma system.

Relativistic plasma wave modes, presenting accelerating behavior, have been recently introduced in Refs.~\cite{lili,faz3,otro}. 
For the case of self-modulation of a circularly polarized electromagnetic or Alfv\'en wave in a weakly or strongly magnetized relativistic
electron-positron plasma with finite temperature, the nonlinear Schr\"odinger
equation that model such effect has been found to be
\cite{faz1,faz2}
\begin{equation}
    i\frac{\partial a}{\partial t}+P \frac{\partial^2 a}{\partial z^2}+  Q|a|^2
a=0\, ,
\label{eq1}
\end{equation}
where $a=a(t,z)$ is the time and space dependent  complex modulational 
amplitude of the circularly polarized electromagnetic wave, under the approximation of a 
slowly time-varying modulation \cite{faz1,faz2}. Here,
$P=c^2/(2\omega)$, where $c$ is the speed of light, and $\omega$ is the frequency of the wave. Differently, $Q$  depend on the limit case of magnetization of the plasma. In the weakly magnetized case  we have that \cite{faz1} 
\begin{equation}
    Q=\frac{3\lambda \omega_p^2\Omega_c^2}{\omega^3 f^5}\, ,
    \label{qq1}
\end{equation}
where $\lambda= e^2/m^2c^4$ (with the electron charge $e$ and mass $m$), $\omega_p$ is the plasma frequency of the electron-positron plasma, $\Omega_c$ is is the cyclotron frequency, and $f$ is the  
thermodynamic function relating the plasma density enthalpy   per unit of mass and unit of number density (being function of temperature). On the other hand, for the strongly magnetized plasma case  we find that \cite{faz2}  
\begin{eqnarray}
  Q=\frac{f\lambda \omega_p^2\omega^3}{4\Omega_c^4}\, .
   \label{qqs}
\end{eqnarray}

It is clear that
 is the background magnetic field, through the cyclotron frequency, the physical quantity that induces the nonlinear behavior of the waves.

\section{Non-accelerating soliton}

Eq.~\eqref{eq1} is usually solved in terms of a  soliton amplitude
propagating with 
 constant speed $v$. This solution can be found by requiring that the amplitude of the electromagnetic wave has the form $|a(t,z)|=|a(z-v t)|$. Straightforwardly, it is found that such solution has the form of a soliton \cite{faz1,faz2}
\begin{eqnarray}
    a(t,z)&=&\,  {\mbox{sech}}\left( \sqrt{\frac{Q}{2 P}}(z- v t)\right) \nonumber\\
    &&\, \exp\left(i \frac{ v }{2P}z-i\left(\frac{v^2}{4P}-\frac{Q}{2}\right) t  \right)\, .
\end{eqnarray}

\section{Accelerating solution}

However, Eq.~\eqref{eq1} can be also solved for an accelerating wave, i.e, we look for solution with amplitude in the form $|a(t,z)|=|a(z+v t-\beta t^2/2)|$, where where $v$ and $\beta$ play the role of a initial velocity and the acceleration of the propagation.

Similar to the case of previous section, this kind of solution can be studied by assuming the form of the plasma wavepacket as
\begin{equation}
    a(t,z)= f\left( \xi \right)\exp\left(i  \eta(t,z)\right)\, ,
    \label{formacc}
\end{equation}
where $f$ is a function to be determined through Eq.~\eqref{eq1}, and depending on the accelerated argument
\begin{eqnarray}
    \xi=\sqrt{\frac{Q}{2P}}\left(z+vt- \frac{\beta}{2} t^2\right) \, .
    \label{argument}
\end{eqnarray}
Notice that we have assumed, in analogy with the constant velocity soliton solution, the same factor $\sqrt{Q/2P}$ for the argument. Also, the phase function $\eta(t,z)$ must be determined. 

Using \eqref{formacc} in 
Eq.~\eqref{eq1} we can determine that when the acceleration is
\begin{equation}
    \beta=Q\sqrt{\frac{Q P}{2}}\, ,
    \label{acceleration}
\end{equation}
and the phase is
\begin{eqnarray}
    \eta(t,z)&=&\frac{Q}{2}\sqrt{\frac{Q}{2P}}\,\left(t\, z-\frac{v}{Q}\sqrt{\frac{2}{QP}}z-\frac{v^2}{Q^2}\sqrt{\frac{Q}{2P}}t\right.\nonumber\\
    &&\left.\qquad \qquad +v\, t^2-\frac{Q}{3} \sqrt{\frac{QP}{2}}  t^3\right)\, ,
\end{eqnarray}
then, the function $f$  fulfills a modified second Painlev\'e equation \cite{clarkP}
\begin{equation}
    \frac{d^2 f}{d \xi^2}-\xi f+2 f^3=0\, .
\label{eq1b}
\end{equation}

Solutions of this equation can be generally studied numerically, but they are no longer solitons. For the current case, they describe an accelerated propagation of a electromagnetic plasma wave train, with non-constant amplitude, and with acceleration $\beta=\sqrt{Q^3 P/2}$ along its direction of propagation. 
The acceleration of this nonlinear plasma wave depends on the magnetization of the plasma through $Q$, given  in Eqs~\eqref{qq1} and  \eqref{qqs}. We emphasize that this accelerated behavior for the plasma is only possible in the
weakly and strongly magnetized limits.

\begin{figure}
{\includegraphics[width = 3.3in]{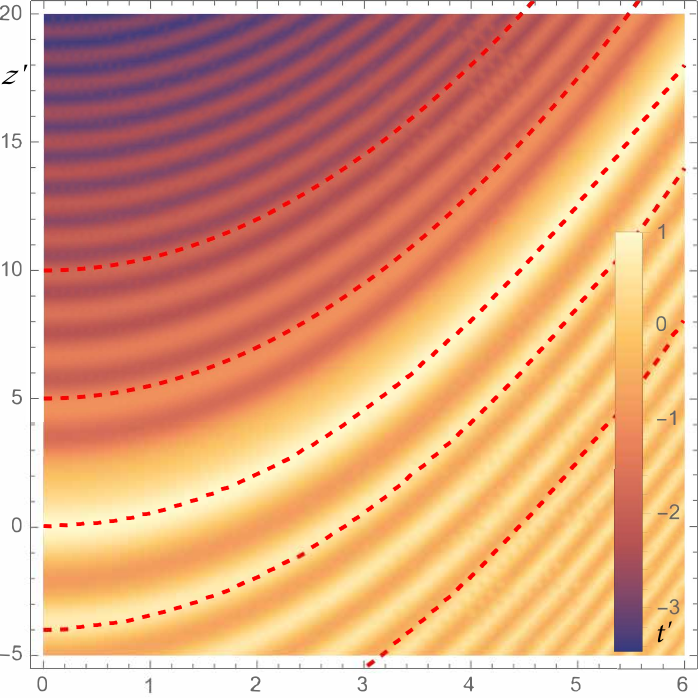}}
\caption{Density plot for $f(\xi)$, with $f(\xi=0)=1$ and 
$d_\xi f(\xi)=0$, in terms of time $t'$ and  distance $z'$, for $v=0$. Red dashed lines corresponds to parabolic trajectories $z'- t'^2/2=\xi_0$, with $\xi_0=-10, -4, \, 0,\, 5,\, 10$.}
\label{figura1}
\end{figure}

In order to show the accelerating behavior of this nonlinear plasma solution, we display in Fig.~\ref{figura1} the density plot for the numerical solution of Eq.~\eqref{eq1b} for $f(\xi=0)=1$,  
$d_\xi f(\xi)=0$, and with $v=0$. This plot shows the magnitude of function $f$ in the $t-z$ space, in terms of normalized time $t'=Q t/\sqrt{2}$, and normalized distance $z'=\sqrt{Q/2P}\, z$. The solution shows curved (parabolic) trajectories for any part of the wave (maxima or minima). This can be explicitly seen through the red dashed lines, that are used as examples. Those lines correspond to $\xi=z'- t'^2/2=\xi_0$, for $\xi_0=-10,-4, \, 0,\, 5,\, 10$. All the red dashed parabolic curves coincide with the dynamics of the plasma wave.
Therefore, we conclude that the whole plasma wave propagates with
acceleration $\beta$ given by \eqref{acceleration}.

\begin{figure}
{\includegraphics[width = 3.3in]{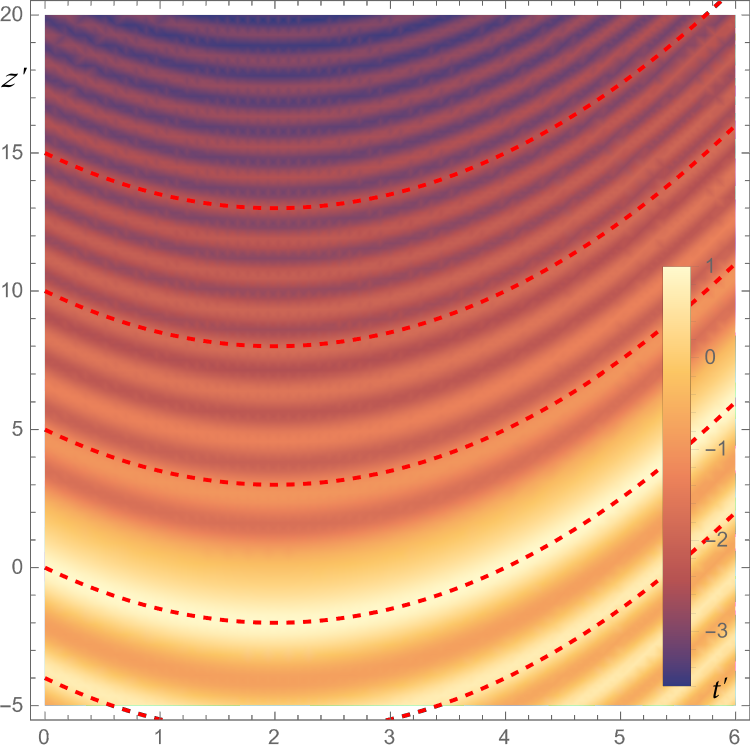}}
\caption{Density plot for $f(\xi)$, with $f(\xi=0)=1$ and 
$d_\xi f(\xi)=0$, in terms of time $t'$,  distance $z'$,  and normalized velocity $v'=2$. Red dashed lines corresponds to parabolic trajectories $z'+ v't'- t'^2/2=\xi_0$, with $\xi_0= -4, \, 0,\, 5,\, 10, \, 15$.}
\label{figura2}
\end{figure}

In Fig.~\ref{figura2} we show what occur when the initial velocity $v$ is considered. For the normalized case $v'=v/\sqrt{P Q}$=2, we display the numerical behavior of $f$ in a density plot in terms of the normalized variables $t'$ and $z'$. For this case, because its accelerating nature, the whole wavepacket change its initial direction of propagation. This occur for any part of the wave, as it is shown with the  parabolic curved in $t - z$ space (in red dashed lines), for  $\xi=z'+ v't'- t'^2/2=\xi_0$, with $\xi_0= -4, \, 0,\, 5,\, 10, \, 15$. The acceleration, therefore, allow to this nonlinear plasma solution to reverse its propagation direction, in a time equal to
\begin{equation}
    t=\frac{v}{\beta}\, .
\end{equation}

Furthermore, although a numerical study of the whole solution is possible, analytical properties of the solution for $f$ can  be found 
in the case when  $\xi\rightarrow -\infty$. In this limit, the solution of the modified second Painlev\'e equation \eqref{eq1b}
behaves as \cite{clarkP}
\begin{equation}
    f(\xi)\approx \kappa\,  {\mbox{Ai}}(\xi)\, ,
\end{equation}
where ${\mbox{Ai}}$ is the Airy function, and $\kappa$ is an arbitrary constant. As the accelerating propagating properties of this electromagnetic plasma wave are present for any value of $\xi$, this solution  pertains to the same family of other accelerating Airy solutions already found in optics and plasmas \cite{Mahalov,lili, otro, faz3,neomi,Jiang,abdo,bouch,esat,panag,moya,Baumgartl,Nikolaos,chong, Kaminer}.

\section{Final remark}

We have presented a new nonlinear plasma solution with accelerating properties. As the initial velocity of the argument \eqref{argument}
is arbitrary, a whole set of new kind of different propagation can be obtained. This is achieved in a relativistic plasma regime, depending on how magnetized the plasma is.

Finally, the  Painlev\'e equation has been explored in different realms of plasma physics \cite{rogers,Sachin,Khater,Ibrahim1,Ibrahim2}. Therefore, this work contributes to show that the  Painlev\'e equation is also a straightforward  consequence of accelerating solutions in relativistic nonlinear plasmas.

\begin{acknowledgements}
FAA thanks to FONDECYT grant No. 1230094 that partially supported this work.
 \end{acknowledgements}

\end{document}